\documentclass[prd,superscriptaddress,amssymb,amsmath,amsfonts,aps,altaffilletter,nofootinbib,letterpaper]{revtex4}
\usepackage{subfigure}
\usepackage[pdftex]{graphicx}

\begin{document}

\begin{abstract}{\begin{center}\textbf{ABSTRACT}\end{center}}
We summarize the sensitivity achieved by the LIGO and Virgo gravitational wave detectors for low-mass compact binary coalescence (CBC) searches during LIGO's sixth science run and Virgo's second and third science runs.  We present strain noise power spectral densities (PSDs) which are representative of the typical performance achieved by the detectors in these science runs. The data presented here and in the accompanying web-accessible data files \cite{DCCpage} are intended to be released to the public as a summary of detector performance for low-mass CBC searches during S6 and VSR2-3.
\end{abstract}

\title{Sensitivity Achieved by the LIGO and Virgo Gravitational Wave Detectors during LIGO's Sixth and Virgo's Second and Third Science Runs}




\affiliation{LIGO - California Institute of Technology, Pasadena, CA  91125, USA$^\ast$}
\affiliation{California State University Fullerton, Fullerton CA 92831 USA$^\ast$}
\affiliation{SUPA, University of Glasgow, Glasgow, G12 8QQ, United Kingdom$^\ast$}
\affiliation{Laboratoire d'Annecy-le-Vieux de Physique des Particules (LAPP), Universit\'e de Savoie, CNRS/IN2P3, F-74941 Annecy-Le-Vieux, France$^\dagger$}
\affiliation{INFN, Sezione di Napoli $^a$; Universit\`a di Napoli 'Federico II'$^b$ Complesso Universitario di Monte S.Angelo, I-80126 Napoli; Universit\`a di Salerno, Fisciano, I-84084 Salerno$^c$, Italy$^\dagger$}
\affiliation{LIGO - Livingston Observatory, Livingston, LA  70754, USA$^\ast$}
\affiliation{Albert-Einstein-Institut, Max-Planck-Institut f\"ur Gravitationsphysik, D-30167 Hannover, Germany$^\ast$}
\affiliation{Leibniz Universit\"at Hannover, D-30167 Hannover, Germany$^\ast$}
\affiliation{Nikhef, Science Park, Amsterdam, the Netherlands$^a$; VU University Amsterdam, De Boelelaan 1081, 1081 HV Amsterdam, the Netherlands$^b$$^\dagger$}
\affiliation{University of Wisconsin--Milwaukee, Milwaukee, WI  53201, USA$^\ast$}
\affiliation{Stanford University, Stanford, CA  94305, USA$^\ast$}
\affiliation{University of Florida, Gainesville, FL  32611, USA$^\ast$}
\affiliation{Louisiana State University, Baton Rouge, LA  70803, USA$^\ast$}
\affiliation{University of Birmingham, Birmingham, B15 2TT, United Kingdom$^\ast$}
\affiliation{INFN, Sezione di Roma$^a$; Universit\`a 'La Sapienza'$^b$, I-00185 Roma, Italy$^\dagger$}
\affiliation{LIGO - Hanford Observatory, Richland, WA  99352, USA$^\ast$}
\affiliation{Albert-Einstein-Institut, Max-Planck-Institut f\"ur Gravitationsphysik, D-14476 Golm, Germany$^\ast$}
\affiliation{Montana State University, Bozeman, MT 59717, USA$^\ast$}
\affiliation{European Gravitational Observatory (EGO), I-56021 Cascina (PI), Italy$^\dagger$}
\affiliation{Syracuse University, Syracuse, NY  13244, USA$^\ast$}
\affiliation{University of Western Australia, Crawley, WA 6009, Australia$^\ast$}
\affiliation{LIGO - Massachusetts Institute of Technology, Cambridge, MA 02139, USA$^\ast$}
\affiliation{Laboratoire AstroParticule et Cosmologie (APC) Universit\'e Paris Diderot, CNRS: IN2P3, CEA: DSM/IRFU, Observatoire de Paris, 10 rue A.Domon et L.Duquet, 75013 Paris - France$^\dagger$}
\affiliation{Columbia University, New York, NY  10027, USA$^\ast$}
\affiliation{INFN, Sezione di Pisa$^a$; Universit\`a di Pisa$^b$; I-56127 Pisa; Universit\`a di Siena, I-53100 Siena$^c$, Italy$^\dagger$}
\affiliation{The University of Texas at Brownsville and Texas Southmost College, Brownsville, TX  78520, USA$^\ast$}
\affiliation{San Jose State University, San Jose, CA 95192, USA$^\ast$}
\affiliation{Moscow State University, Moscow, 119992, Russia$^\ast$}
\affiliation{LAL, Universit\'e Paris-Sud, IN2P3/CNRS, F-91898 Orsay$^a$; ESPCI, CNRS,  F-75005 Paris$^b$, France$^\dagger$}
\affiliation{NASA/Goddard Space Flight Center, Greenbelt, MD  20771, USA$^\ast$}
\affiliation{The Pennsylvania State University, University Park, PA  16802, USA$^\ast$}
\affiliation{Universit\'e Nice-Sophia-Antipolis, CNRS, Observatoire de la C\^ote d'Azur, F-06304 Nice$^a$; Institut de Physique de Rennes, CNRS, Universit\'e de Rennes 1, 35042 Rennes$^b$, France$^\dagger$}
\affiliation{Laboratoire des Mat\'eriaux Avanc\'es (LMA), IN2P3/CNRS, F-69622 Villeurbanne, Lyon, France$^\dagger$}
\affiliation{Washington State University, Pullman, WA 99164, USA$^\ast$}
\affiliation{INFN, Sezione di Perugia$^a$; Universit\`a di Perugia$^b$, I-06123 Perugia,Italy$^\dagger$}
\affiliation{INFN, Sezione di Firenze, I-50019 Sesto Fiorentino$^a$; Universit\`a degli Studi di Urbino 'Carlo Bo', I-61029 Urbino$^b$, Italy$^\dagger$}
\affiliation{University of Oregon, Eugene, OR  97403, USA$^\ast$}
\affiliation{Laboratoire Kastler Brossel, ENS, CNRS, UPMC, Universit\'e Pierre et Marie Curie, 4 Place Jussieu, F-75005 Paris, France$^\dagger$}
\affiliation{Rutherford Appleton Laboratory, HSIC, Chilton, Didcot, Oxon OX11 0QX United Kingdom$^\ast$}
\affiliation{IM-PAN 00-956 Warsaw$^a$; Astronomical Observatory Warsaw University 00-478 Warsaw$^b$; CAMK-PAN 00-716 Warsaw$^c$; Bia{\l}ystok University 15-424 Bia{\l}ystok$^d$; IPJ 05-400 \'Swierk-Otwock$^e$; Institute of Astronomy 65-265 Zielona G\'ora$^f$,  Poland$^\dagger$}
\affiliation{University of Maryland, College Park, MD 20742 USA$^\ast$}
\affiliation{University of Massachusetts - Amherst, Amherst, MA 01003, USA$^\ast$}
\affiliation{Canadian Institute for Theoretical Astrophysics, University of Toronto, Toronto, Ontario, M5S 3H8, Canada$^\ast$}
\affiliation{Tsinghua University, Beijing 100084 China$^\ast$}
\affiliation{University of Michigan, Ann Arbor, MI  48109, USA$^\ast$}
\affiliation{The University of Mississippi, University, MS 38677, USA$^\ast$}
\affiliation{Charles Sturt University, Wagga Wagga, NSW 2678, Australia$^\ast$}
\affiliation{Caltech-CaRT, Pasadena, CA  91125, USA$^\ast$}
\affiliation{INFN, Sezione di Genova;  I-16146  Genova, Italy$^\dagger$}
\affiliation{Pusan National University, Busan 609-735, Korea$^\ast$}
\affiliation{Carleton College, Northfield, MN  55057, USA$^\ast$}
\affiliation{Australian National University, Canberra, ACT 0200, Australia$^\ast$}
\affiliation{The University of Melbourne, Parkville, VIC 3010, Australia$^\ast$}
\affiliation{Cardiff University, Cardiff, CF24 3AA, United Kingdom$^\ast$}
\affiliation{INFN, Sezione di Roma Tor Vergata$^a$; Universit\`a di Roma Tor Vergata, I-00133 Roma$^b$; Universit\`a dell'Aquila, I-67100 L'Aquila$^c$, Italy$^\dagger$}
\affiliation{University of Salerno, I-84084 Fisciano (Salerno), Italy and INFN (Sezione di Napoli), Italy$^\dagger$}
\affiliation{The University of Sheffield, Sheffield S10 2TN, United Kingdom$^\ast$}
\affiliation{RMKI, H-1121 Budapest, Konkoly Thege Mikl\'os \'ut 29-33, Hungary$^\ast$}
\affiliation{INFN, Gruppo Collegato di Trento$^a$ and Universit\`a di Trento$^b$,  I-38050 Povo, Trento, Italy;   INFN, Sezione di Padova$^c$ and Universit\`a di Padova$^d$, I-35131 Padova, Italy$^\dagger$}
\affiliation{Inter-University Centre for Astronomy and Astrophysics, Pune - 411007, India$^\ast$}
\affiliation{University of Minnesota, Minneapolis, MN 55455, USA$^\ast$}
\affiliation{California Institute of Technology, Pasadena, CA  91125, USA$^\ast$}
\affiliation{Northwestern University, Evanston, IL  60208, USA$^\ast$}
\affiliation{The University of Texas at Austin, Austin, TX 78712, USA$^\ast$}
\affiliation{E\"otv\"os Lor\'and University, Budapest, 1117 Hungary$^\ast$}
\affiliation{University of Adelaide, Adelaide, SA 5005, Australia$^\ast$}
\affiliation{University of Szeged, 6720 Szeged, D\'om t\'er 9, Hungary$^\ast$}
\affiliation{Embry-Riddle Aeronautical University, Prescott, AZ   86301 USA$^\ast$}
\affiliation{National Institute for Mathematical Sciences, Daejeon 305-390, Korea$^\ast$}
\affiliation{Perimeter Institute for Theoretical Physics, Ontario, Canada, N2L 2Y5$^\ast$}
\affiliation{National Astronomical Observatory of Japan, Tokyo  181-8588, Japan$^\ast$}
\affiliation{Universitat de les Illes Balears, E-07122 Palma de Mallorca, Spain$^\ast$}
\affiliation{Korea Institute of Science and Technology Information, Daejeon 305-806, Korea$^\ast$}
\affiliation{University of Southampton, Southampton, SO17 1BJ, United Kingdom$^\ast$}
\affiliation{Institute of Applied Physics, Nizhny Novgorod, 603950, Russia$^\ast$}
\affiliation{Lund Observatory, Box 43, SE-221 00, Lund, Sweden$^\ast$}
\affiliation{Hanyang University, Seoul 133-791, Korea$^\ast$}
\affiliation{Seoul National University, Seoul 151-742, Korea$^\ast$}
\affiliation{University of Strathclyde, Glasgow, G1 1XQ, United Kingdom$^\ast$}
\affiliation{Southern University and A\&M College, Baton Rouge, LA  70813, USA$^\ast$}
\affiliation{University of Rochester, Rochester, NY  14627, USA$^\ast$}
\affiliation{Rochester Institute of Technology, Rochester, NY  14623, USA$^\ast$}
\affiliation{Hobart and William Smith Colleges, Geneva, NY  14456, USA$^\ast$}
\affiliation{University of Sannio at Benevento, I-82100 Benevento, Italy and INFN (Sezione di Napoli), Italy$^\ast$}
\affiliation{Louisiana Tech University, Ruston, LA  71272, USA$^\ast$}
\affiliation{McNeese State University, Lake Charles, LA 70609 USA$^\ast$}
\affiliation{University of Washington, Seattle, WA, 98195-4290, USA$^\ast$}
\affiliation{Andrews University, Berrien Springs, MI 49104 USA$^\ast$}
\affiliation{Trinity University, San Antonio, TX  78212, USA$^\ast$}
\affiliation{Southeastern Louisiana University, Hammond, LA  70402, USA$^\ast$}
\author{J.~Abadie$^\text{1}$}\noaffiliation\author{B.~P.~Abbott$^\text{1}$}\noaffiliation\author{R.~Abbott$^\text{1}$}\noaffiliation\author{T.~D.~Abbott$^\text{2}$}\noaffiliation\author{M.~Abernathy$^\text{3}$}\noaffiliation\author{T.~Accadia$^\text{4}$}\noaffiliation\author{F.~Acernese$^\text{5a,5c}$}\noaffiliation\author{C.~Adams$^\text{6}$}\noaffiliation\author{R.~Adhikari$^\text{1}$}\noaffiliation\author{C.~Affeldt$^\text{7,8}$}\noaffiliation
% ...

\collaboration{$^\ast$The LIGO Scientific Collaboration and $^\dagger$The Virgo Collaboration}
\noaffiliation




\maketitle

\section{Introduction}

The LIGO and Virgo gravitational wave observatories recently completed their final joint science run until advanced detectors come online. The observation was LIGO's sixth science run and Virgo's second and third science runs. LIGO operated two instruments, the four kilometer Hanford detector (H1) and the four kilometer Livingston detector (L1) from July 07, 2009 to October 20, 2010. The two kilometer Hanford interferometer, which was included in the S5/VSR1 CBC analyses, was decommissioned prior to the start of this run and did not collect any data. The three kilometer Virgo interferometer (V1) operated during roughly the same timespan, with however a major commissioning break between its second and third science runs, from Jan 11, 2010 to August 6, 2010.

In this note, we summarize the sensitivity achieved by these detectors during the latest runs from the perspective of low-mass (inspiral-only) CBC searches. The strain noise power spectral density (PSD) is a complete characterization of the sensitivity of a detector but generally only meaningful over timescales of about an hour. Over longer time scales, the noise characteristics of the detectors typically vary significantly, for instance as the morning traffic picks up. More importantly, the detectors themselves change over time as they undergo weekly maintenance and occasional larger scale upgrades. The notion of PSD does not immediately have meaning when applied to the detector performance over the entire duration of the run. To solve this problem, we choose a single ``representative'' block of time for each detector and compute the PSD for each detector in this block. We take the resulting PSD as representative of the typical sensitivity to gravitational waves from CBCs achieved by the detectors in S6/VSR2-3.

We choose the respresentative time for each detector by looking at its inspiral horizon distance distribution and selecting a time for which the detector operated with a horizon distance close the the mode of this distribution. The inspiral horizon distance is equal to the largest distance at which an equal-mass compact binary inspiral would accumulate an average SNR of 8 in the detector. The inspiral horizon distance contains less information but is derived from the PSD and includes information specific to CBC signals. In this sense, the horizon distance is a useful measure of the sensitivity of a detector to gravitational waves from CBCs at a given time. In this article, we gather the inspiral horizon distance data generated during S6/VSR2-3 low-mass CBC search \cite{S6lowmass} and use the results to identify PSDs that are representative of detector performance for CBC searches during these science runs.

The plots and data presented here are intended to be released to the public as a summary of detector performance for CBC searches during S6 and VSR2-3 \cite{DCCpage}.  These results use exactly the same science segments and analysis code \cite{TmpltBank} that was used in the low-mass CBC search in S6 and VSR2-3. In the next section, we review how the inspiral horizon distance is computed from the PSD. In section 3, we present the inspiral horizon distance data obtained from S6/VSR2-3 CBC analyses. We then use these data to identify PSDs which are representative of the detector sensitivity to inspirals for this science run. We close this note with a discussion of one potential pitfall in using these spectra.

\section{Inspiral Horizon Distance}

In LIGO and Virgo data analysis applications, we treat the strain noise in a detector as a stationary random process.  If the noise in the detector were truly stationary, then the noise spectral density would completely characterize the sensitivity of the detector as a function of frequency. The power spectral density $S_n(f)$ for a stationary random process $n(t)$ is defined implicitly by the relation
\begin{eqnarray}
\label{psd}\frac{1}{2} S_n(f)\delta(f-f') =  \langle \tilde{n}(f)\tilde{n}^*(f') \rangle ,
\end{eqnarray}
where $\tilde{n}(f)$ is the Fourier transform of the random process.  The spectral density is a measure of the mean square noise fluctuations at a given frequency.   
As mentioned above, the noise in the LIGO and Virgo detectors is not stationary.  However, by measuring the spectral density over a short enough timescale, we are able to approximate the noise as stationary.  The chosen timescale must also be long enough that we can form an accurate estimate of the spectral density.  In the S6/VSR2-3 CBC searches, the spectral density was computed on 2048-second blocks of contiguous data \cite{FindChirp}.  We account for long timescale non-stationarities by using a different spectral density for every 2048 seconds.

In assessing the overall performance of a detector for CBC searches, we use the inspiral horizon distance data from S6 and VSR2-3 to identify the ``typical'' sensitivity of the interferometers.  The inspiral horizon distance of a detector is the distance at which an optimally oriented and optimally located equal-mass compact binary inspiral would give an average signal to noise ratio (SNR) of $\rho=8$ in the interferometer. If $\tilde{h}(f)$ represents the Fourier transform of the expected signal, then the average SNR this signal would attain in a detector with spectral density $S_n(f)$ is given by
\begin{eqnarray}
\label{general_ir} \langle \rho \rangle= \sqrt{4 \int_0^\infty \frac{|\tilde{h}(f)|^2}{S_n(f)} df}.
\end{eqnarray}
We find the inspiral horizon distance by setting $\langle \rho \rangle = 8$ and solving for the distance $D$ to the inspiral event which parametrizes the waveform $\tilde{h}(f)$.  Thus, the inspiral horizon distance combines the spectral density curve with the expected inspiral waveform to produce a single quantity that summarizes the sensitivity of the detector at a given time.

Practical considerations require modifications to the limits of the integral.  In the CBC search code, we compute the signal to noise ratio by
\begin{eqnarray}
\label{range} \langle \rho \rangle= \sqrt{4 \int_{f_{low}}^{f_{high}} \frac{|\tilde{h}(f)|^2}{S_n(f)} df}.
\end{eqnarray}
The lower limit is determined by our ability to characterize the noise at low frequencies.  In the S6 CBC search, we took $f_{low}=40$Hz as the low frequency cut-off in computing the inspiral horizon distance.  For Virgo in VSR2-3, the low frequency cut-off was $f_{low}=50$Hz.  The upper limit of the integral is the innermost stable circular orbit (ISCO) frequency,
\begin{eqnarray}
f_{isco} = \frac{c^3}{6\sqrt{6}\pi G M},
\end{eqnarray}
where $M$ is the total mass of the binary system.  For binary neutron star (BNS) systems, for which we take $m=1.4M_\odot$, $f_{isco}= 1570$Hz.

The inspiral waveform for CBCs is accurately given in the frequency domain by the stationary phase approximation.  For an optimally oriented and optimally located equal mass binary, the signal that appears at the interferometer (in this approximation) is given by
\begin{eqnarray}
\label{spa}
\tilde{h}(f) =  \frac{1}{D}\left(\frac{5\pi }{24c^3}\right)^{1/2}(G\mathcal{M})^{5/6}(\pi f)^{-7/6} e^{i\Psi(f;M)},
\end{eqnarray}
where $\mathcal{M} = \mu^{3/5}M^{2/5}$ is the chirp mass of the binary, $D$ is the distance to the binary and $\Psi$ is a real function of $f$, parametrized by the total mass $M$.  Setting $\langle \rho \rangle = 8$ and inserting this waveform into eqn. \ref{range}, we find that the inspiral horizon distance is given by
\begin{eqnarray}
\label{range0} D = \frac{1}{8}\left(\frac{5\pi }{24c^3}\right)^{1/2}(G\mathcal{M})^{5/6}\pi^{-7/6} \sqrt{4 \int_{f_{low}}^{f_{high}} \frac{f^{-7/3}}{S_n(f)}df }.
\end{eqnarray}
The inspiral horizon distance is defined for optimally located and oriented sources. To compare with previous results, note that the sensitive range of an interferometer gravitational-wave detector was considered by Finn and Chernoff \cite{Finn:1993}, though here we following the conventions of Allen, {\it et. al.} \cite{FindChirp} and Brown \cite{DBrownThesis}. Furthermore, if we divide the inspiral horizon distance given here by 2.26 we obtain the SenseMon range \cite{Sutton} reported as a figure of merit in the LIGO and Virgo control rooms, where the factor
of 2.26 comes from averaging over a uniform distribution of source sky locations and orientations.

In practice, it is convenient to measure distances in Mpc and  mass in $M_\odot$. It is useful therefore to specialize eqn. \ref{range0} to this unit system.  Further, since we measure the strain $h(t)$ at discrete time intervals $\Delta t = 1/f_s$, the spectral density is only known with a frequency resolution of $\Delta f = f_{s}/N$, where $N$ is the number of data points used to measure $S_n(f)$.  By putting $f=k/(N\Delta t)$ into eqn. \ref{range0} and grouping terms by units, we arrive at the expression
\begin{eqnarray}
 D \approx \frac{1}{8} \mathcal{T} \sqrt{\frac{4\Delta t}{N} \sum_{k_{low}}^{k=k_{high}} \frac{(k/N)^{-7/3}}{S_n(k)} } \mathrm{ Mpc },
\end{eqnarray}
where
\begin{eqnarray}
\mathcal{T} = \left( \frac{5}{24\pi^{4/3}} \right)^{1/2}
              \left( \frac{ \mu }{ M_\odot } \right)^{1/2}
              \left( \frac{ M }{ M_\odot } \right)^{1/3}
              \left( \frac{ G M_{\odot}/c^2 }{\mathrm{1 Mpc}} \right)
              \left( \frac{ G M_\odot/c^3 }{\Delta t } \right)^{-1/6},
\end{eqnarray}
for the inspiral horizon distance in Mpc. Since it is convenient to work with the binary system's component masses, we have also replaced the chirp mass $\mathcal{M}$ with the reduced mass $\mu$ and the total mass $M$.  Written this way, the inspiral horizon distance in Mpc is easily computed from the binary component masses in $M_\odot$.

\begin{figure}
\includegraphics[scale=0.4]{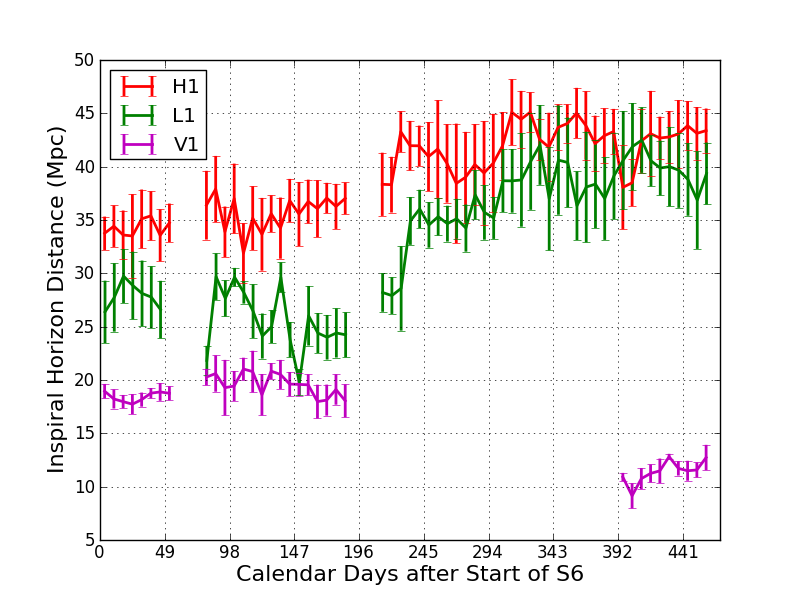}
\includegraphics[scale=0.4]{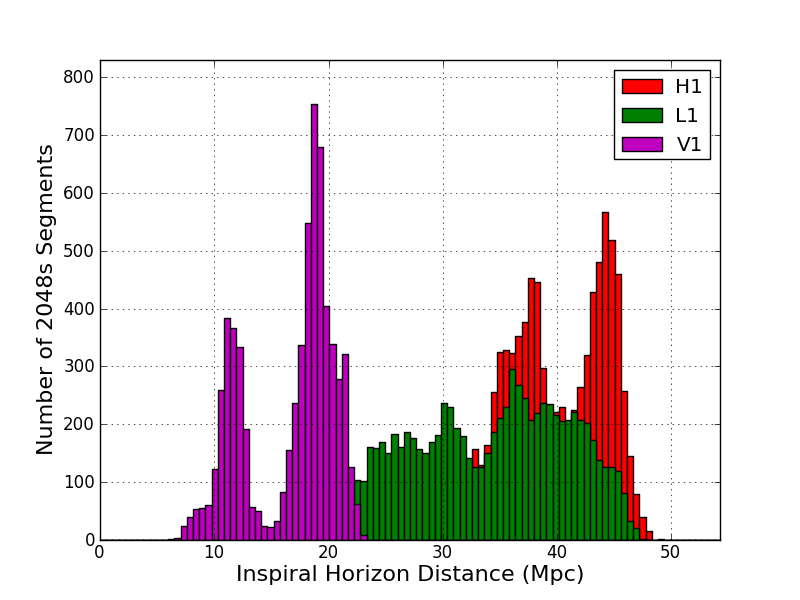}
\caption{(a) Inspiral horizon distance as a function of time during S6-VSR2/3.  The average inspiral horizon distances for each week in S6 and VSR2-3.  As an indication of the weekly variations, we have included error bars corresponding to the standard deviation of the inspiral horizon distance during each week. (b) Distribution of 1.4-1.4 solar mass inspiral horizon distance for the three gravitational wave detectors H1, L1, and V1 for the joint LIGO-Virgo science run consisting of S6 and VSR2/3.  The histogrammed data consists of the same 2048-second analyzed segments from the S6 and VSR2/3 CBC searches.}
\label{rangevtime}
\end{figure}

\section{Summary of Inspiral Horizon Distance Data}

Here we present the horizon distance data collected from the final data products produced during the S6/VSR2-3 lowmass CBC search \cite{S6lowmass}. We have collected the data, rather than computing the inspiral horizon distance directly, in order to ensure that we analyze the exact same science segments and use the exact same analysis code as used in the LIGO/Virgo CBC searches.

In Fig. \ref{rangevtime}a, we plot the average BNS inspiral horizon distance for each of the three detectors as a function of time. We use a window of one week and the points on the plot correspond to the average inspiral horizon distance for all science segments beginning in that week. The error bars attached to the points indicate the standard deviation in the inspiral horizon over the course of the given week. This figure highlights the variability in sensitivity throughout the run and the reason it is difficult to identify a single time for each detector with a typical or average sensitivity. In Fig. \ref{rangevtime}b, we histogram the BNS inspiral horizon distance for the three detectors H1, L1, and V1. The bimodal behavior seen in the LIGO and Virgo detectors is largely due to a significant commissioning break in S6 and commissioning in Virgo between VSR2 and VSR3. These commissioning breaks will be described in detail in a later publication on the S6/VSR2-3 runs.

In the actual S6/VSR2-3 CBC analysis, the inspiral horizon is computed for ($n$)-($n$) solar mass binaries for $n$ an integer. Previous documents \cite{Nada} however have plotted the horizon distance for the canonical $1.4-1.4$ solar mass binary neutron star. In order to simplify comparison to previous results, we rescale the obtained distributions by $(2.8/2)^{5/6}$ corresponding to the ratio of chirp masses of a $1.0-1.0$ solar mass system and a $1.4-1.4$ solar mass system.  This scaling ignores the fact that $f_{isco}$ is different for the two mass pairs, but this is negligible since the signal template is buried in the noise at such high frequencies.

\begin{figure}
\includegraphics[scale=0.65]{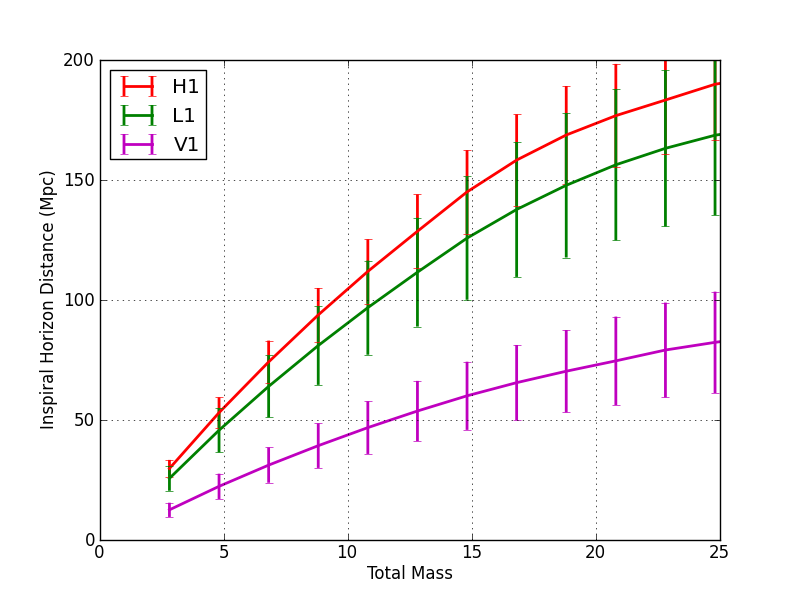}
\caption{Mean inspiral horizon distance as a function of mass for the three gravitational wave detectors H1, L1 and V1 during S6-VSR2/3.  The error bars on the curves extend from one standard deviation below to one standard deviation above the mean.}
\label{rangevmass}
\end{figure}

In Fig. \ref{rangevmass}, we show the mean inspiral horizon distance for each interferometer as a function of the binary total mass, assuming equal mass binaries.  This plot reflects the mean performance of the detector over various frequency bands.  As the component mass becomes higher, the upper cutoff frequency $f_{high}=f_{isco}$ becomes smaller and smaller.  This means that the inspiral horizon distance focuses on a narrower band around the lower cutoff $f_{low}=40$Hz (or $f_{low}=50$Hz in the case of Virgo).  The inspiral horizon distance takes into account only the inspiral stage of the CBC event, while for high-mass systems ($M> 25M_{\odot}$) the merger and ringdown stages of the coalescence occur in the LIGO and Virgo sensitive band.  For total masses greater than 25$M_\odot$, the inspiral-only range begins to fall over, which is not indicative of the sensitivity of the detector for these systems. For these binary systems, we use Effective One Body Numerical Relativity (EOBNR) waveform templates that include the merger and ringdown stages and our sensitivity is significantly improved relative to an inspiral-only analysis \cite{S5highmass}.

\section{Representative Power Spectral Density}

\begin{figure}
\includegraphics[scale=0.65]{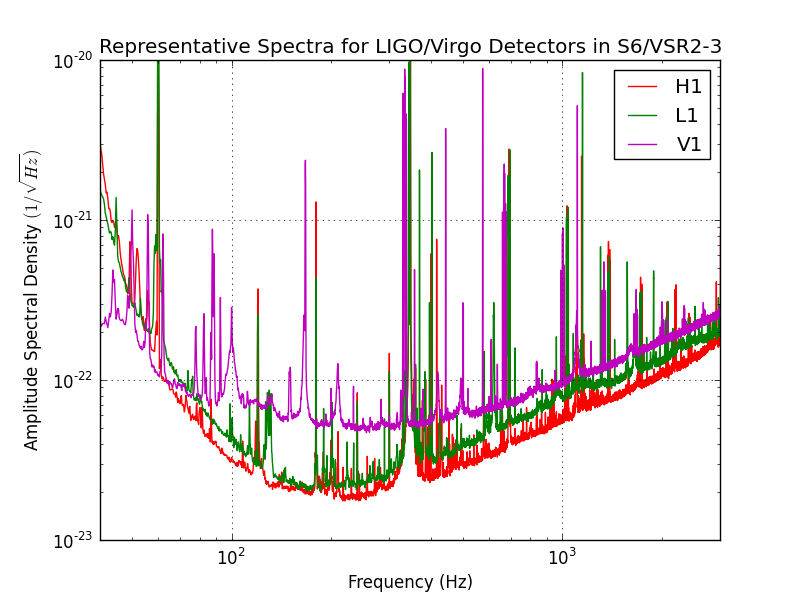}
\caption{Representative spectral density curves for LIGO and Virgo detectors during S6 and VSR2-3.  We plot here the amplitude spectral density, which is the square root of the power spectral density, since the strength of a gravitational wave signal is proportional to the strain induced in the interferometer and the sensitivty of a detector is therefore most naturally expressed in terms of this quantity. These spectral density curves correspond to May 9, 2010 (GPS 957474700) for H1, February 27, 2010 (GPS 951280082) for L1 and  August 30, 2009 (GPS 935662133) for V1.  These times are chosen such that the inspiral horizon distance for each detector at that time coincides with the mode of its inspiral horizon distance distribution, as given by the midpoint of the most populated bin in Fig. \ref{rangevtime}b.}
\label{psdcurves}
\end{figure}

In Fig. \ref{psdcurves}, we give representative spectral density curves for each of the three detectors during S6 and VSR2-3.  The chosen representative curve corresponds to a time when the detector operated near the mode of its inspiral horizon distance distribution shown in Fig. \ref{rangevtime}b.  The algorithm used to compute the spectral densities is described in detail in \cite{FindChirp}.  The parameters needed in order to reconstruct our results are given in Table \ref{params}.  The first column in Table \ref{params} gives a list of parameter names and symbols, which are the same names and symbols used in \cite{FindChirp}.  The second and third columns gives the values of these parameters used in S6/VSR2-3 CBC searches.  These parameters can be used to reproduce the inspiral horizon distance data accompanying this note.  The fourth column gives the values of the parameters used to compute the representative spectral density curves shown here.  In making our choice of parameters for computing representative spectra, we sacrificed frequency resolution ($\Delta f = 1/T$) for PSD accuracy (which increases with $N_S$).

\begin{table}
\begin{center}
\caption{Parameters used in the computation of the spectral density.}
\label{params}
$ $\\
\begin{tabular}{l|c|c|c}
FINDCHIRP parameter \cite{FindChirp}  & S6 low-mass & VSR2-3 low-mass & representative spectra\\
\hline
sample rate ($1/\Delta t$) & 4096 Hz & 4096 Hz & 16384 Hz\\
data block duration ($T_{block}$) & 2048s & 2048s &  2048s\\
number of data segments ($N_S$) & 15 & 15 & 1023\\
data segment duration ($T$) & 256s & 256s & 4s\\
stride ($\Delta$) & 524288 & 524288 & 32768
\end{tabular}
\end{center}
\end{table}

\begin{figure}
\includegraphics[scale=0.65]{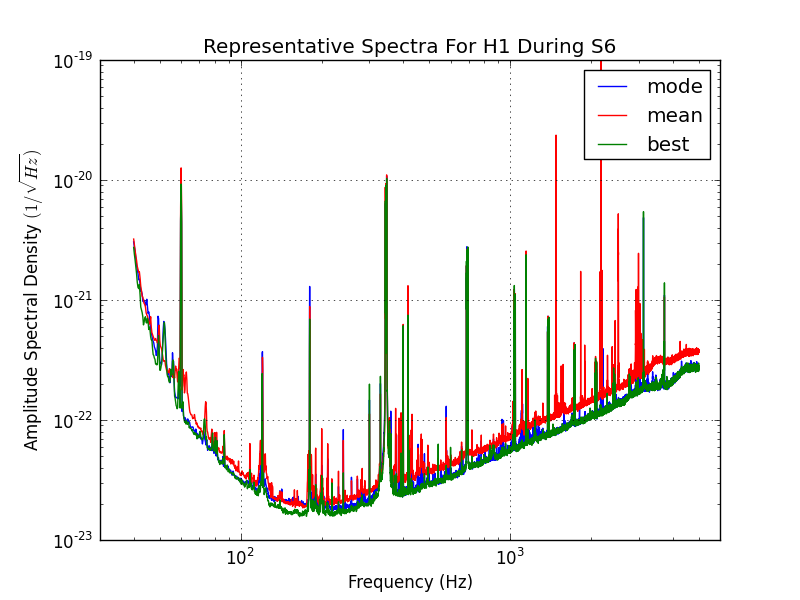}
\caption{Various possibilities for a representative PSD for H1 during S6. These spectral density curves correspond to times when the detector operated near its S6 mode (44.1Mpc), mean (39.5Mpc) and maximum (49.3Mpc) inspiral horizon distance.  The times chosen are May 9, 2010 (GPS 957474700) for the mode, November 4, 2010 (GPS 941365351) for the mean and July 4, 2010 (GPS 962268343) for the maximum.}
\label{h1curves}
\end{figure}

One potential pitfall with using these spectra is that the choice of representative PSD for a detector is not obvious. Here we illustrate the degree to which our choice of using the mode affected the chosen PSD. We compare the spectra for H1 corresponding to times when H1 operated near its mode to times when it operated near its mean and maximum of its inspiral horizon distance distribution. In Table \ref{stats}, we provide a quantitative summary of the low-mass inspiral horizon distance distributions for a 1.4--1.4$M_\odot$ binary given in Mpc.  We see that the horizon distance varies by roughly 10\% between its mode and mean.  This suggests that spectral density curves from a detector's most common sensitivity (mode) may differ significantly from the spectral density of a detector's ``average'' performance. To illustrate this point, we plot in Fig. \ref{h1curves} three spectra for H1 from different times in S6.


\begin{table}
\begin{center}
\caption{Inspiral Horizon Distance Summary for a 1.4--1.4$M_\odot$ Binary}
\label{stats}
$ $\\
\begin{tabular}{l|cccc}
           & H1 & L1 & V1\\
\hline
mean & 39.5 & 34.0 & 16.6 \\
max & 49.3 & 47.2 & 23.2 \\
mode & 44.1 & 36.6 & 18.8 \\
std & 4.7 & 6.9 & 3.9 
\end{tabular}
\end{center}
\end{table}

All of the data used here have been computed using the final version of calibration used in the CBC searches. Note that the noise spectra presented here are subject to systematic uncertainties associated with the strain calibration. These uncertainties can be up to $\pm$15\% in amplitude. For more detail, see references \cite{Goetz:2009,VirgoCalibration}.

\section*{Acknowledgements}

The authors gratefully acknowledge the support of the United States National Science Foundation for the construction and operation of the LIGO Laboratory, the Science and Technology Facilities Council of the United Kingdom, the Max-Planck-Society, and the State of Niedersachsen/Germany for support of the construction and operation of the GEO600 detector, and the Italian Istituto Nazionale di Fisica Nucleare and the French Centre National de la Recherche Scientiﬁque for the construction and operation of the Virgo detector. The authors also gratefully acknowledge the support of the research by these agencies and by the Australian Research Council, the Council of Scientiﬁc and Industrial Research of India, the Istituto Nazionale di Fisica Nucleare of Italy, the Spanish Ministerio de Educaci´on y Ciencia, the Conselleria d’Economia Hisenda i Innovaci´o of the Govern de les Illes Balears, the Foundation for Fundamental Research on Matter supported by the Netherlands Organisation for Scientiﬁc Research, the Royal Society, the Scottish Funding Council, the Polish Ministry of Science and Higher Education, the FOCUS Programme of Foundation for Polish Science, the Scottish Universities Physics Alliance, The National Aeronautics and Space Administration, the Carnegie Trust, the Leverhulme Trust, the David and Lucile Packard Foundation, the Research Corporation, and the Alfred P. Sloan Foundation. This document has been assigned LIGO Document No. LIGO-T1100338.

Anyone using the information in this document and associated material (S6/VSR2/VSR3 noise spectra, inspiral ranges, observation times) in a publication or talk must acknowledge the US National Science Foundation, the LIGO Scientific Collaboration, and the Virgo Collaboration.  Data files associated with the results and plots presented in this document can be found here: https://dcc.ligo.org/cgi-bin/public/DocDB/ShowDocument?docid=63432.  Please direct all questions to the corresponding author (Stephen Privitera, stephen.privitera@ligo.org).  Please inform the corresponding author and the LSC and Virgo spokespeople (LSC-Spokesperson@ligo.org and Jean-Yves.Vinet@oca.eu, respectively) if you intend to use this information in a publication.

\clearpage

\bibliography{S6_detector_sensitivity}{}
\bibliographystyle{unsrt}


\end{document}